\documentclass{svmult}
\usepackage{graphicx,amsfonts,amsmath,amssymb,bbm} 
\usepackage{dsfont}
\usepackage{mathrsfs}

\newcommand{\cF}{\mathcal{F}}

\newcommand{\cR}{\mathcal{R}}

\begin{document}
\bibliographystyle{spmpsci}
\title*{On Some Open Problems in Many-Electron Theory}
\author{Voler Bach and Luigi Delle Site}
\institute{Volker Bach \at Carl-Friedrich-Gauss-Fakult\"{a}t,  Institut f\"{u}r Analysis und Algebra, Technische Universit\"{a}t Braunschweig, Braunschweig, Germany\\ \email{v.bach@tu-bs.de}\\ Luigi Delle Site \at Institute for Mathematics, Freie Universit\"{a}t Berlin, Arnimallee 6, D-14195 Berlin, Germany\\ \email{luigi.dellesite@fu-berlin.de}}

\maketitle
\section*{What's next?}
Considering the chapters of this book as a road-map for the progress
in the field of many-electron theories along the path of multi-disciplinarity, one can
draw some basic conclusions. It is clear that computational physics
and chemistry are going to have a role of increasing importance 
in the description of many-electron systems, as technological progress leads to ever more
powerful computers. It is also clear that the other disciplines can
certainly help this process assuring the highest efficiency in
developing computational algorithms. However, beyond applications, the
investigation of fundamental questions of many-electron systems is by far
not saturated. This implies that while on one side we will have
increasingly efficient and accurate numerical descriptions of matter,
with system sizes and time scales directly comparable to those of
experiments, on the other side, part of our effort should be devoted
to the analysis of fundamental aspects of the problem which up to now
did not find a satisfactory treatment. Mel Levy and Elliott Lieb are
two of the most prominent researchers who have dedicated their efforts
to the investigation of fundamental questions in many-electron
theory. Their results have not only revolutionized the
theoretical approach of the field, but, directly or indirectly,
allowed for a quantum jump in the computational treatment of realistic
systems as well. For this reason, at the conclusion of this journey
across different disciplines, we have asked Mel Levy and Elliott Lieb
to provide us with a list of open problems, summarized below, which
they believe will be a worth challenge for the future also in the
perspective of a synergy among the various disciplines.\\

\section*{Two open problems formulated by Mel Levy}
\begin{itemize}
\item {\bf Is there a closed form expression for ground-state energy from
  ground-state density?}\\

For the interacting physical system of interest, assume that one is
given the exact ground-state density, $\rho({\bf r})$, associated with
electron-nuclear attraction potential $v({\bf r})$.  Is there a
closed-form expression that gives the ground-state energy, $E_{gs}$,
in terms of $\rho({\bf r})$ and $v({\bf r})$?\\ I am not aware of a
theorem that states that the answer is definitely no because the
system is interacting.  In fact the answer would be yes if each term
in the electron-electron repulsion operator were squared. Then, for
instance, $E_{gs}$ would equal $(1/2) V_{en}$ for an atom, where
$V_{en}$ is the electron-nuclear attraction energy.  This result
follows from use of the virial theorem with the fact that the
electron-electron repulsion operator here exhibits the same
homogeneous coordinate scaling as the kinetic operator.  Perhaps the
answer is also yes for real physical systems, with a more complicated
relation.\\ Presently, there are known bounds for real physical
systems with exact electron-electron repulsion operators. For
instance, it can be shown that:
\begin{equation}
          (1/2) V_{en} <  E_{gs}  <  (1/3) V_{en}
\end{equation}
for any atom, and there exist analogous bounds for any system. For
systems other than atoms, gradients of $v({\bf r})$ or $\rho({\bf r})$
are needed. The left inequality simply comes from using the virial
theorem and neglecting half the electron-electron repulsion. The right
inequality comes from use of a generalized variational theorem where the number
of electrons in the wave function is greater than in the Hamiltonian.\\

\item {\bf Is the Ionization Energy always greater than the Electron
  Affinity for Coulomb systems?}\\

To my knowledge there is no proof that the ionization energy is always
greater than the electron affinity for Coulomb systems, even though
this property is so very important in the study of band gaps and
strong correlations.  In other words, there does not appear to be a
general proof that $I[N+1] < I[N]$ where $I[M]$ is the ionization
energy of the M-electron system.  It should be noted, however, that
the special case $I[2] < I[1]$ has been shown \cite{mel1}. Also, exceptions to
$I[N+1] < I[N]$ have been found for non-Coulomb systems\cite{Lieb1983}.\\ The
property $I[N+1] < I[N]$ can be proven for certain Coulomb atoms.
Consider those atoms for which $I[N+1] < I[N]$ when the
electron-electron repulsion is absent.  For these atoms, It can 
be shown that $I[N+1] < I[N]$ when the full electron-electron
repulsion operator, $V_{ee}$, is present, provided that the nuclear
charge is high enough. This result follows from use of coordinate scaling.
\end{itemize}

\section*{An Open Problem about exchange-correlation energy formulated by Elliott Lieb}
The Lieb-Oxford Inequality \cite{LiebOxford1981}  states that, for {\em
any}
normalized, symmetric or anti-symmetric $N$-particle wave function,
$\Psi_N$,
\begin{align} \label{eq-I.1}
\bigg\langle \Psi_N \; & \bigg| 
\sum_{1 \leq m < n \leq N} \frac{1}{|x_m-x_n|} \, \Psi_N \bigg\rangle
\\[1ex] \nonumber
& \ \geq \
\frac{1}{2} \int \frac{  \rho_\Psi(x) \, \rho_\Psi(y) \: d^3x \, d^3y}{|x-y|}
\; - \; C_{LO} \int \rho_\Psi^{4/3}(x) \: d^3x,
\end{align}
where $C_{LO} = 1.68$ and 
\begin{align} \label{eq-I.2}
\rho_\Psi(x) \ = \ 
N \, \int |\Psi(x, x_2, \ldots, x_N)|^2 \: d^3x_2 \cdots d^3x_N
\end{align}
is the one-particle density corresponding to $\Psi_N$, with $\int
\rho_\Psi(x) \, d^3x = N$. The first term on the right side is called the
{\em direct energy} and the difference between it and the left side of
(\ref{eq-I.1}) is called the {\em exchange- correlation energy} and is
denoted by $E_{xc}$. Thus, (\ref{eq-I.1}) gives a lower bound to $E_{xc}$
and the goal is to improve it. 
Motivated by inequality (\ref{eq-I.1}), much effort
\cite{BenguriaTusek2012,BenguriaGallegoTusek2012,ChanHandy1999,Levyetal1993,PerdewBurke1996,VelaMedeletal2009}
has gone into determining the optimal
value for the constant, which lies
between about $1.45$ and $1.68$. The number 1.45 is based on numerical
values for the charged electron gas in a uniform background. A rigorous
lower value of $1.23$ for $C_{LO}$
was obtained in \cite{LiebOxford1981}. Current opinion is that the true
answer is close to this $1.45$ figure.

The inequality (\ref{eq-I.1}) is also valid for a density {\em matrix} that
is not a pure state, but we will continue to think in terms of $\Psi$. 

On  the other hand, Dirac had shown much earlier \cite{Dirac1930} that for a
{\em free Fermi} gas in its {\em lowest energy state}, the above inequality
would become an asymptotic
equality, for large $N \gg 1$ by replacing $C_{LO}$ by the
spin-dependent constant $C_D  =
0.93 \, q^{-2/3}$, where $q$ is the number of spin states available to the
fermions. Electrons have $q=2$ and in this case $C_D =0.93\cdot 2^{-1/3}
=0.74 \ll 1.68$. 

The open problem discussed here is how to reconcile the Dirac $E_{xc}$ and
the $-C_{LO} \int\rho_\psi^{4/3}$ lower bound for $E_{xc}$ in
(\ref{eq-I.1}). Before presenting it in detail, let us first mention  a
problem that looks similar, but is quite different. 

There are quite a few papers in the literature that try take into
account the variation of $
\rho_\Psi(x)$ with $x$. An especially interesting result is by
Benguria, Bley and Loss \cite{BenguriaLoss2012} who (essentially)  bound
$E_{xc}$
by $-1.45 \int \rho_\Psi^{4/3} - C' \langle
\sqrt{\rho_\Psi } \, \big| \,|p|\, \big| \sqrt{\rho_\Psi }\rangle$ for some
constant $C'$ and
where $|p| = \sqrt{\nabla^2}$. They note that the added term,
$-C'\langle \cdots\rangle$ is often very small compared to first term and,
therefore, the effective constant is approximately $1.45$, as anticipated.

Our  open problem is, however, {\em not}  about the effect on
$E_{xc}$ of the spatial  variation of the quantity $\rho_\psi$, which is,
after all, the 
envelope of the many-body wave function. It refers to the {\em
internal} local  kinetic
energy, which is present even where $\rho_\Psi $ is constant and which is
not visible in $\rho_\psi$. The LO bound 
on $E_{xc}$, whatever the sharp constant $C_{LO}$ might turn out to be, is
relevant only if the underlying wave function has a kinetic energy that 
is much higher than the minimum kinetic energy density for a gas having
particle density $\rho$, which equals 
$$
\tfrac{3}{5} (6 \pi^2)^{2/3}q^{-2/3}\rho_\Psi(x)^{5/3}.
$$
If the kinetic energy is close to this minimum then the Dirac constant
$0.74$
is relevant. If it is far from this minimum then the constant $C_{LO} $ is
relevant. 

We may define the {\em kinetic
energy density} as the positive function 
\begin{align} \label{eq-I.3}
T_\Psi(x) 
\ = \ 
N \, \int \big| \nabla_x \Psi(x, x_2, \ldots, x_N) \big|^2 
\: dx_2 \cdots dx_N ,
\end{align}
Clearly, the total kinetic energy is $T= \int T_\Psi (x) d^3x$,\   just as 
the particle number is $N= \int \rho_\Psi (x) d^3x$. Since quantum mechanics
can be formulated in momentum-space as well as in $x$-space, $T_\Psi (x)$
should be as relevant as $\rho_\Psi(x)$. 

\medskip
\boxed{OPEN PROBLEM:} Find a lower bound to $E_{xc}$ that involves both
$T_\Psi (x)$
and $\rho_\Psi(x)$ by using the ratio                            
$$
\cR_{\Psi}(x) =  \frac{T_\Psi(x)}{  
\tfrac{3}{5} (6 \pi^2)^{2/3}q^{-2/3}\rho_\Psi(x)^{5/3}}
$$
to improve the LO type bound (or some other bound) in a way that (locally)  
reduces to the Dirac $E_{xc}$ when $\cR_\Psi(x) $ is small. When 
$\cR_\Psi(x)  $ is
large the effective constant (locally)  should be $C_{LO}$.\hfill
$\blacksquare$

\end{document}